\documentclass[aps,prl,twocolumn,footinbib,showpacs,floatfix]{revtex4}
\usepackage{amssymb}
\usepackage[pdftex]{graphicx}	
	\DeclareGraphicsExtensions{.pdf,.png,.jpg}
\usepackage[applemac]{inputenc}
\usepackage{amsmath,amssymb,amsthm}
\usepackage{mathbbol,bbm}
\usepackage{graphicx,float}
\usepackage[final]{showkeys}
\usepackage{bm,dsfont}
\usepackage{color}
\usepackage{hyperref}

\definecolor{myblue}{rgb}{0.3,0.4,0.85}
\definecolor{myred}{rgb}{0.8,0.,0.2}
\definecolor{mygreen}{rgb}{0.6,0.8,0.2}

\usepackage{booktabs}

\begin{document}
\title{Chiral Topological Superconductors Enhanced by Long-Range Interactions}
\author{Oscar Viyuela$^{1,2,3}$, Liang Fu$^{1}$, and Miguel Angel Martin-Delgado$^{2}$}
\affiliation{1. Department of Physics, Massachusetts Institute of Technology, Cambridge, MA 02139, USA\\
2. Departamento de F\'{\i}sica Te\'orica I, Universidad Complutense, 28040 Madrid, Spain\\
3. Department of Physics, Harvard University, Cambridge, MA 02318, USA}

\vspace{-3.5cm}

\begin{abstract}
We study the phase diagram and edge states of a two-dimensional p-wave superconductor with long-range hopping and pairing amplitudes. New topological phases and quasiparticles different from the usual short-range model are obtained. When both hopping and pairing terms decay with the same exponent, one of the topological chiral phases with propagating Majorana edge states gets significantly enhanced by long-range couplings. On the other hand, when the long-range pairing amplitude decays more slowly than the hopping, we discover new topological phases where propagating Majorana fermions at each edge pair nonlocally and become gapped even in the thermodynamic limit. Remarkably, these nonlocal edge states are still robust, remain separated from the bulk, and are localized at both edges at the same time. The inclusion of long-range effects is potentially applicable to recent experiments with magnetic impurities and islands in 2D superconductors.
\end{abstract}


%74.20.Mn Non-conventional mechanisms of Superconductivity
%71.10.Pm	Fermions in reduced dimensions (anyons, composite fermions, Luttinger liquid, etc.) (for anyon mechanism in superconductors, see 74.20.Mn)
%03.65.Vf Geometric & Topological Phases
%05.30.Rt Quantum Phase transitions
% 85.25.-j Superconducting Devices.
%74.20.-z	Theories and models of superconducting state
%03.67.Lx	Quantum computation architectures and implementations

\maketitle

%%%%%%%%%%%%%%%%%%%%%%%%%%%%%%%%%%%%%%%%%%%%%%%%%%%%%%%%%%%%%%%%

\noindent {\it 1. Introduction.---} Topological superconductors are novel quantum phases of matter \cite{rmp1,rmp2,LibroBernevig} with unconventional pairing symmetries that host gapless Majorana zero modes. These are exotic quasiparticles that can realize non-Abelian statistics \cite{Read_et_al00} unlike bosons or fermions. Remarkably, their intrinsic topological protection and robustness against disorder make them ideal candidates as building blocks in quantum computation \cite{rmp3,rmp4,Alicea_et_al_11,Baranov_et_al_13} and quantum information processing \cite{Kitaev01,Mazza_et_al13}.  

As first shown in \cite{Fu_Kane_2008}, the surface state of a 3D topological insulator can host Majorana zero modes by proximity coupling a conventional s-wave superconductor. Topological superconductivity can also be induced by replacing the topological insulator with a semiconductor with strong spin-orbit interaction \cite{Sau_et_al_2010,Sau2_et_al_2010,Alicea_2010,Sau3_et_al_2010,Oreg_et_al_2010}. Indeed, several experiments in proximity-induced superconducting nanowires \cite{Mourik_et_al12,Deng_et_al12,Das_et_al12,Albrecht_et_al16} and topological insulator-superconductor heterostructures \cite{Wang_et_al12,He_et_al14,Sun_et_al16,He_et_al17} have shown signatures for the presence of Majorana zero modes and chiral edge states. In all these platforms, the superconducting pairing amplitude is of purely short-range nature.

An alternative approach to realizing chiral p-wave Hamiltonians is to deposit magnetic atoms on top of a conventional s-wave superconductor substrate \cite{Nadj_et_al14,Pawlak_et_al15}. For a periodic 1D array of these magnetic impurities \cite{Pascual_et_al16,Ruby_et_al17}, a p-wave Hamiltonian with intrinsic long-range pairing \cite{Nadj_et_al13,Pientka2013,Klinovaja_et_al13,Pientka2014,Li_et_al_16,Kaladzhyan_et_al16} is induced, provided that the length of the chain is small compared to the coherence length of the host superconductor \cite{Pientka2013}. Novel topological phase diagrams \cite{Viyuela_et_al16,Pachos_16,Luca_16,Alecce_17} have been analyzed for certain 1D Hamiltonians with long-range interactions \cite{Niui_12,DeGottardi_13,Vodola_et_al14,Gong2015_1,Gong2015_2,Tudela_15,Vodola_et_al16,Vodola_et_al17,Dutta_17}. Interestingly, massless Majorana end modes were shown to pair up into a topological massive Dirac fermion \cite{Viyuela_et_al16} localized simultaneously at the two ends of a 1D long-range p-wave superconductor. More recently, spatially extended Shiba states \cite{Yu_65,Shiba_68,Rusinov_69} of magnetic impurities in 2D superconductors have been experimentally observed \cite{Menard_et_al15,Menard_et_al16,Heinrich_et_al17}. When they conform to planar arrays  \cite{Ronty_et_al_15,Li_et_al_16_2D,Kaladzhyan_et_al16_2}, long-range couplings between the impurity states emerge, raising the question of whether new topological phases in 2D long-range systems can be found.

%%%%%%%%%%
\begin{figure}[t]
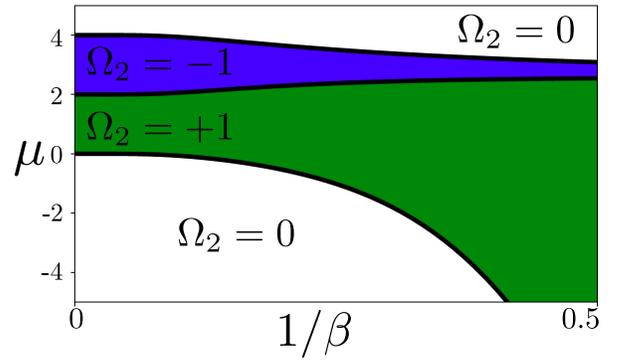

\centering
\includegraphics[width=0.9\columnwidth]{{{Fig1_v3}}}%   
\caption{Topological phase diagram for the case of equal long-range decay $\alpha=\beta$, as function of the chemical potential $\mu$ and the inverse of the decaying exponent $\beta$. At $1/\beta\rightarrow0$, we recover the phase diagram for the short-range p-wave topological superconductor. Using the topological winding number $\Omega_2$, we identify a trivial phase ($\Omega_2=0$), a topological phase with a certain chirality ($\Omega_2=+1$) that gets enlarged, and a topological phase with opposite chirality ($\Omega_2=-1$) that gets suppressed for very strong long-range decay.} 
\label{fig_phasediagram}
\end{figure}
%%%%%%%%%%%%

Motivated by previous experiments with magnetic atoms \cite{Nadj_et_al14,Pascual_et_al16,Menard_et_al15,Heinrich_et_al17} and islands \cite{Menard_et_al16}, we compute the topological phase diagram (see Fig.~\ref{fig_phasediagram}) and the edge states of a 2D p-wave superconductor with long-range hopping and pairing couplings. We find topological phases and quasiparticles that were absent in the usual short-range model \cite{Read_et_al00,LibroBernevig}. When the power-law decay for hopping and for pairing amplitudes is the same, we observe different phenomena. (A) Enhanced Topological Chiral Phase (ETCP): a topological chiral phase with stable propagating Majorana edge states even for very strong long-range effects. (B) Suppressed Topological Chiral Phase (STCP): the topological phase with opposite chirality is slowly diminished and even disappears completely in the infinite-range limit. (C) Chiral Symmetry Inversion (CSI): a topological phase transition between the two chiral phases is driven when tuning the decaying exponent. On the other hand, when the pairing amplitude decays more slowly than the hopping, there is a critical value of the decaying exponent at which propagating edge states become gapped. Remarkably, these states remain chiral and localized at the edges, which leads to intrinsic protection against backscattering.

%%%%%%%
%%%%%%%
\noindent {\it 2. Long-range chiral topological superconductor.---} We consider a 2D lattice of spinless fermions with long-range hopping and pairing amplitudes,
\begin{align}
H=&\sideset{}{'}\sum_{\substack{m,l\\r,s}}\Big(-\frac{t}{d^{\beta}_{r,s}}c^{\dagger}_{l+r,m+s}c_{l,m} + \frac{\tilde{\Delta}(r,s)}{d^{\alpha}_{r,s}}c^{\dagger}_{l+r,m+s}c^{\dagger}_{l,m}+\nonumber\\
&+\text{h.c.}\Big)-(\mu-4t)\sum_{l,m}c^{\dagger}_{l,m}c_{l,m},
\label{Hlr}
\end{align}	
%\end{equation}
where the sum $\sideset{}{'}\sum_{r,s}$ runs over all integers $r$ and $s$ except $r=s=0$. The function $d_{r,s}=\sqrt{r^2+s^2}$ is the Euclidean distance between fermions and $\mu$ is the chemical potential, with a certain origin offset for convenience. The hopping amplitude $t$ decay with exponent $\beta$ of the distance $d_{r,s}$, and the pairing amplitude 
\begin{equation}
\tilde{\Delta}(r,s)=\frac{\Delta}{d_{r,s}}(r + {\rm i}s),
\label{Delta}
\end{equation}
decay with exponent $\alpha$. Without loss of generality, we fix $\Delta=t=1/2$ in the rest of the paper. Eq.~\eqref{Delta} is the most natural extension of a p-wave symmetry when considering long-range effects. Moreover, Hamiltonian \eqref{Hlr} belongs to the D symmetry class of topological insulators and superconductors \cite{Ludwig,Kitaev_2009}, whatever $\alpha$ and $\beta$. Fluctuation effects over the mean field Hamiltonian have not been considered in the present work.

\begin{figure}[t]
\centering
\includegraphics[width=\columnwidth]{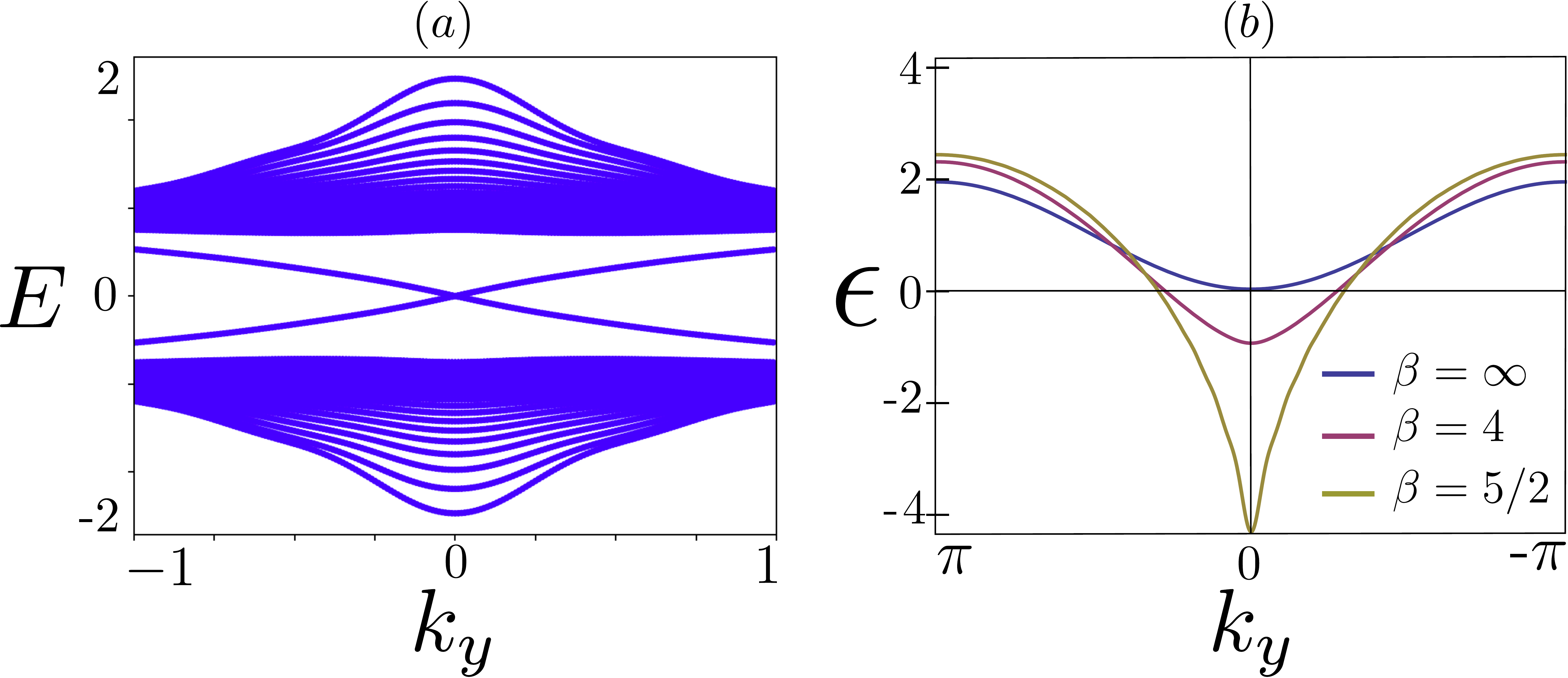}
\caption{(a) Energy dispersion relation for cylindrical boundary conditions (periodic along the $y-$direction) and long-range hopping and pairing decaying with the same exponent $\alpha=\beta=\frac{5}{2}$. We set $\mu=1$, corresponding to the topological phase $\Omega_2=+1$. We identify two energy bands corresponding to bulk states and two propagating edge states crossing at $k_y=0$. (b) Kinetic energy per particle $\epsilon(k_y)=g_{\beta}(k_x=0,k_y)-2$ as a function of $k_y$. The broader the energy range, the larger the chiral sector with $\Omega=+1$. The blue line corresponds to short-range $\beta=\infty$, the red line to $\beta=4$, the yellow one to $\beta=\frac{5}{2}$. We can see a pronounced peak at the $\Gamma$ point for strong long-range effects.}
\label{fig_pairhop}
\end{figure}

At large values of the decaying exponents ($\alpha,\beta \rightarrow \infty$), Eq.~\eqref{Hlr} has a well-defined limit and we recover a short-range chiral p-wave Hamiltonian \cite{Volovik_99,Read_et_al00,Ivanov_01,LibroBernevig}. The system hosts vortices with non-Abelian anyonic statistics \cite{Ivanov_01} that are of great relevance in proposals for topological quantum computation \cite{rmp3}. In addition, propagating Majorana modes appear at the boundary, displaying a linear dispersion at low momentum that connects the particle and hole bands (see Fig.~\ref{fig_pairhop}a). 

Now we analyze the fate of these topologically protected edge states when we add long-range hopping and pairing amplitudes as in Eq.~\eqref{Hlr}. To this end, we assume periodic boundary conditions in the $y-$direction, where the Fourier-transformed operators are
\begin{equation}
c_{(l,m)}=\frac{1}{\sqrt{L}}\sum_{k_y\in {\rm B.Z.}}{\rm e}^{{\rm i}k_ym}c_{l, k_y},
\end{equation}
with $L$ the number of sites along the $y-$direction. Thus, we can rewrite Hamiltonian \eqref{Hlr} as $H=\sum_{k_y} H(k_y)$, where
\begin{align}
&H(k_y)=\sum_{l} \Bigg[ \sum_r \Big(\Gamma_r^{\beta}(k_y)c^{\dagger}_{l+r,k_y}c_{l,k_y} + \label{Hky}\\ 
&+\gamma_r^{\alpha}(k_y)c^{\dagger}_{l+r,k_y}c^{\dagger}_{l,-k_y} + \text{h.c.}\Big) - (\mu-4t) c^{\dagger}_{l,k_y}c_{l,k_y} \Bigg]\nonumber,
\end{align}

with 
\begin{equation}
\Gamma_r^{\beta}(k_y) = -t \sum_{s=0}^{L} \frac{{\rm e}^{-{\rm i} k_ys}}{d_{r,s}^{\beta}}~~~\text{and}~~~\gamma_r^{\alpha}(k_y) = \sum_{s=0}^{L} \frac{\tilde{\Delta}(r,s)~{\rm e}^{-{\rm i} k_ys}}{d_{r,s}^{\alpha}}.
\label{gammas}
\end{equation}

Since we assume periodic boundary conditions, the definition of the distance function $d_{r,s}$ and the pairing function $\tilde{\Delta}(r,s)$ need to be slightly modified by replacing $s~\rightarrow L-s$ whenever $s>L/2$. We are now in a position to study the edge state structure of Hamiltonian \eqref{Hky}. We consider two different cases: i) $\alpha=\beta$ where both hopping and pairing amplitudes decay with the same exponent, and ii) $\alpha<\beta$ where the pairing term is longer range than the hopping.

%%%%%%%%%%
%%%%%%%%%%%%

%%%%%%%
%%%%%%%
\noindent {\it 3. Chiral Majorana edge states.---} Let us first consider the case where long-range hopping and pairing terms decay with the same exponent $\alpha=\beta$. The motivation comes from the experimental and theoretical results regarding magnetic atoms on top of s-wave topological superconductors both in one-dimensional chains \cite{Nadj_et_al13,Pientka2013,Pientka2014,Li_et_al_16,Nadj_et_al14} and planar structures \cite{Ronty_et_al_15,Li_et_al_16_2D,Menard_et_al15,Heinrich_et_al17}, where long-range effects have been shown to play an even more important role than in 1D systems. 

In what follows, we show the link between the edge state physics and the different topological phases. We demonstrate that the topological phase diagram gets modified by the effect of long-range couplings (see Fig.~\ref{fig_phasediagram}), creating an imbalance between different chiral phases. We assume periodic boundary conditions in both spatial directions, where the Hamiltonian \eqref{Hlr} takes the form $H=\sum_{\bf k}\Psi^{\dagger}_{\bf k} H({\bf k}) \Psi_{\bf k}$ in the Nambu-spinor basis of paired fermions with $H_{\bf k}=E_{\bf k}{\bf n}({\bf k})\cdot {\bm \sigma}$, where $\bm{\sigma}=(\sigma_x,\sigma_y,\sigma_z)$ are the Pauli matrices and
\begin{eqnarray}
E_{\bf k}&=& \sqrt{\Big(f_{\alpha}^2({\bf k})+h_{\alpha}^2({\bf k})\Big)+ \Big( g_{\beta}({\bf k}) + \mu-2 \Big)^2},\label{Ek}\\
{\bf n}({\bf k})&=&\frac{-1}{E_{\bf k}}\Big(f_{\alpha}({\bf k}), h_{\alpha}({\bf k}), g_{\beta}({\bf k}) + \mu-2 \Big).
\label{nk}
\end{eqnarray}
Here, $E_{\bf k}$ is the energy dispersion relation and ${\bf n}({\bf k})$ is the so-called winding vector, expressed in terms of series of trigonometric functions
\begin{eqnarray}
f_{\alpha}({\bf k})&=& \sum_{\substack{r,s}} \frac{s}{d_{r,s}^{\alpha+1}}\sin{(k_xr + k_ys)},\\
h_{\alpha}({\bf k})&=& \sum_{\substack{r,s}} \frac{r}{d_{r,s}^{\alpha+1}}\sin{(k_xr + k_ys)},\\
g_{\beta}({\bf k})&=& \sum_{\substack{r,s}} \frac{\cos{(k_xr + k_ys)}}{d_{r,s}^{\beta}},
\label{fk}
\end{eqnarray}
where the sum $\sum_{r,s}$ runs over all integers $r$ and $s$ except $r=s=0$, and $d_{r,s}$ is the Euclidean distance on the ${\cal T}^2$-torus.

We define a winding number $\Omega_2$ in terms of the winding vector ${\bf n}({\bf k})$, which reads as 
\begin{equation}\label{w1}
\Omega_2=\frac{1}{2\pi}\int_{\rm BZ}{\bf n}({\bf k}).\Big(\partial_{k_x}{\bf n}({\bf k}) \times \partial_{k_y}{\bf n}({\bf k})\Big) d^2\bm{k},
\end{equation}
where BZ stands for Brillouin zone. The winding number is an integer topological invariant that characterizes a continuous mapping between the crystalline momentum ${\bf k} \in T^2$ and the winding vector ${\bf n}({\bf k}) \in S^2$. This is also called the Pontryagin index. The topological transition points $\mu_c^\beta$ can be identified out of the gap-closing points of $E_{\bf k}$ given in Eq.~\eqref{Ek},
\begin{eqnarray}
\mu_{c1}^{\beta}&=&2-g_{\beta}(k_x=0,k_y=0),\nonumber\\
\mu_{c2}^{\beta}&=&2-g_{\beta}(k_x=0,k_y=\pi),\label{muc}\\
\mu_{c3}^{\beta}&=&2-g_{\beta}(k_x=\pi,k_y=\pi).\nonumber
\end{eqnarray}

%%%%%%%%%%%%
%%%%%%%%%%%%
\begin{figure*}[t]
\includegraphics[width=\textwidth]{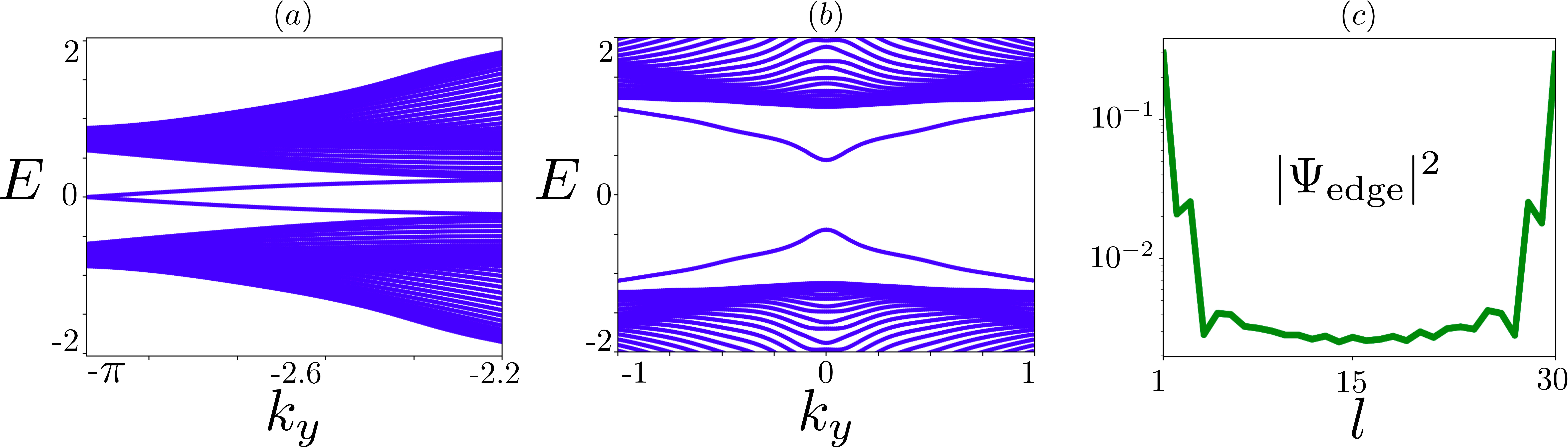}
\caption{Energy dispersion relation $E(k_y)$ and wave function probability for cylindrical boundary conditions and $\alpha<\beta$. We fix $\alpha=1.7$, $\beta=8$. (a) For $\mu=3$ we identify two energy bands corresponding to bulk states and two propagating edge states with a certain chirality crossing at $k_y=\pi$. (b) For $\mu=1$ the edge states of opposite chirality are now gapped. (c) We plot the wave function probability at $k_y=0$ for the gapped edge states showing edge localization.}
\label{LR_pairing}   
\end{figure*}
%%%%%%%%%%%%%%%%%%%%%%%%%%
%%%%%%%%%%%%%%%%%%%%%%%%%%

As shown in Fig.~\ref{fig_phasediagram}, for $\beta\gg1$, the topological phase diagram coincides with the short-range chiral p-wave Hamiltonian \cite{Read_et_al00,LibroBernevig}, with $\mu_{c1}^{\beta=\infty}=0$, $\mu_{c2}^{\beta=\infty}=2$, $\mu_{c1}^{\beta=\infty}=4$. Thus, for $0<\mu<2$, the Majorana modes have a definite chirality ($\Omega_2=+1$), whereas for $2<\mu<4$ they have the opposite ($\Omega_2=-1$). We now discuss the consequences of adding long-range effects that significantly change the boundaries and the phases of the short-range case. If the chemical potential $\mu<\mu_{c1}^{\beta}$ or $\mu>\mu_{c3}^{\beta}$, then $\Omega_2=0$ and the system is in a trivial phase with no edge states. However, when $\mu_{c1}^{\beta}<\mu<\mu_{c2}^{\beta}$, then $\Omega_2=+1$ and there are propagating edge states with definite chirality. As shown in Fig.~\ref{fig_pairhop}a the edge state at each boundary cross at $k_y=0$ when connecting the particle and hole bands. Interestingly enough, this topological phase gets significantly enlarged for $\mu<0$, since the divergence of the function $g_{\beta}(k_x=0,k_y=0)$ at the $\Gamma$ point shifts the location of the transition point $\mu_{c1}^{\beta}$ towards $-\infty$ as the system becomes long-ranged, leading to the ETCP mechanism. This is of great relevance for experimental realizations of p-wave superconductors and Majorana edge states without the need of a precise fine-tuning of the chemical potential. Physically, the effect is explained due to a broadening of the quasiparticle kinetic energy with respect to the short-range case, as shown in Fig.~\ref{fig_pairhop}b. As we decrease the long-range exponent $\beta$, the energy peak at the $\Gamma$ point is more pronounced; thus, the range of chemical potential at which topological superconductivity occurs gets enlarged.  

A distinct topological phase can be found if $\mu_{c2}^{\beta}<\mu<\mu_{c3}^{\beta}$. This phase is characterized by a winding number $\Omega_2=-1$ (see Fig.~\ref{fig_phasediagram}) and the presence of propagating edge states with opposite chirality. As shown in Fig.~\ref{LR_pairing}a the edge states at each boundary cross at $k_y=\pi$. Surprisingly, this topological phase gets progressively suppressed when the long-range terms become important (STCP). This creates an effective asymmetry between the two chiral sectors, whose mathematical explanation is two-fold. First, in the short-range limit ($\beta\gg1$), the two sectors $\mu>2$ and $\mu<2$ of Hamiltonian \eqref{Hlr} are connected through the unitary transformation $c_{i,j} \rightarrow (-1)^{i+j} c_{i,j}^\dagger$ that inverts the chirality. However, the long-range terms break this symmetry leading to an asymmetry between the two sectors. The second reason is that $g_{\beta}({\bf k})$ is an alternating and convergent series at the high-symmetry points $X$ and $M$; thus, according to Eq.~\eqref{muc}, the critical points $\mu_{c2}^{\beta}$ and $\mu_{c3}^{\beta}$ always remain finite. In fact, in the infinite range limit $\beta=0$, they take on the same value $\mu_{c2}^{\beta=0}=\mu_{c3}^{\beta=0}$ and the topological phase involving the edge crossing at $k_y=\pi$ ends disappearing, as shown in Fig.~\ref{fig_phasediagram}. Finally, we can induce topological phase transitions between the two different chiral phases by tuning the decaying exponent at a fixed chemical potential. This mechanism, dubbed chiral symmetry inversion (CSI) is a distinct feature of these long-range topological systems \cite{remark}.
We notice that for $\alpha>\beta$ the pairing amplitude decays faster than the hopping, thus, the edge states never become gapped. Moreover, the phase boundaries in Fig. 1 only depend on the exponent $\beta$ (see Eq. (12)), and therefore would remain unaltered.

%$X$ $(k_x=0,k_y=\pi), (k_x=\pi,k_y=0)$ and $M$ $(k_x=\pi,k_y=\pi)$

%%%%%%%
%%%%%%%
\noindent {\it 4. Non-local gapped edge states.---} We now consider the situation where the long-range pairing amplitude decays slower than the hopping, i.e., $\alpha<\beta$. This situation could be achieved by periodic driving of a topological superconductor, as shown in \cite{Benito_et_al14} for the 1D Kitaev chain \cite{Kitaev01}. We demonstrate that depending on the pairing exponent $\alpha$ and the chirality of the topological phase, the Majorana edge states become gapped despite remaining localized and protected at the boundary. 

Without loss of generality, we assume $\beta\gg1$. First, we focus on the topological phase where $0<\mu<2$ and Chern number $\Omega_2=+1$. We have previously shown in Fig.~\ref{fig_pairhop}a that there are propagating edge states crossing at $k_y=0$, provided $\alpha=\beta$. But the situation is different for $\alpha<\beta$. By means of finite-size-scaling, we numerically show that when $\alpha\lesssim2$ the chiral edge states crossing at $k_y=0$ become gapped due to purely long-range effects as depicted in Fig. \ref{LR_pairing}b. This result coincides with a divergent group velocity ${\bf v}_g=\nabla_{\bf k} E({\bf k})$ at the $\Gamma$ point (with periodic boundary conditions), signaling strong long-range effects within this phase. In Fig. \ref{LR_pairing}c, we plot the spatial distribution $|\psi(x)|^2$ of the edge-states wave function for the $k_y=0$ mode proving edge localization.

The edge localization in the $x-$direction, and the bulk-edge gap depicted in Fig. \ref{LR_pairing} are robust against random fluctuations of the chemical potential in the $x-$direction. Disordered impurities in the $y-$direction could lead to elastic scattering between states with momentum $k_y$ and $-k_y$, respectively, which implies backscattering and localization of the propagating edge states. However, the chiral nature of the massive edge states protects them against backscattering since the modes at $k_y$ and $-k_y$ are localized at different edges. It is only at very low momentum $k_y\approx0$ that the two modes at the edges hybridize, as shown in Fig. \ref{LR_pairing}c of the manuscript. This is a difference with respect to the short-range case. However, this hybridization happens only very close to $k_y=0$, thus, the density of states that are subject to disorder effects is extremely small. Moreover, at large momentum $k_y\gtrsim\frac{\pi}{2}$, the situation is quite the opposite. For strong long-range effects, the edge states remain localized at each edge and do not hybridize, whereas for the short-range model the edge states merge into bulk states. Hence, at large momentum our new edge states are more robust against disorder and back-scattering than the propagating Majorana modes in the short-range model. In addition, these states are topologically protected by a non-trivial Chern number $\Omega_2=+1$.

On the other hand, if $\alpha\gtrsim2$ the Majorna edge states are stable. Second, we analyze the phase with $2<\mu<4$ and Chern number $\Omega_2=-1$. We find propagating edge states crossing at $k_y=\pi$, no matter the exponent $\alpha$. We note that at $k_y=\pi$ the group velocity is not divergent even for very strong long-range; thus, the edge-crossing point is protected as shown in Fig.~\ref{LR_pairing}a. In summary, we have proven that the long-range pairing induces a gap in the edge states depending on the chirality of the topological phase. It is also interesting to analyze the trivial region $\mu<0$, where in the short-range limit $\alpha\gg1$ there are no edge states. However, we numerically find that below $\alpha=5/2$ protected gapped edge states are formed out of originally trivial bulk states. These are now localized at the two edges at the same time, being non-local. The energy dispersion relation and the edge localization are similar to Fig.~\ref{LR_pairing}b and Fig.~\ref{LR_pairing}c respectively.

%%%%%%%%
%%%%%%%%
\noindent {\it 5. Conclusions.---} We found that long-range hopping and pairing amplitudes deeply modify the topological phase diagram of a two-dimensional p-wave topological superconductor. We have shown that when the decay exponents for the pairing and the hopping amplitudes are equal $\alpha=\beta$, one of the topological phases is greatly enhanced (ECTP), while the topological phase with opposite chirality is slowly suppressed (SCTP) for very strong long-range effects (see Fig~\ref{fig_phasediagram}). By tuning the decaying exponent $\beta$, we can induce topological phase transitions between the two different chiral phases (CSI) at fixed chemical potential $\mu$. On the other hand, if $\alpha<\beta$, i.e., the pairing decays slower than the hopping amplitude, rich phenomena occur depending on the exponent $\alpha$ and the chemical potential $\mu$: i) propagating edge states are topologically stable whatever the exponent $\alpha$; ii) Majorana modes become gapped, though remaining localized at the edges and protected against backscattering; iii) previously trivial bulk states give rise to protected gapped edge states. The long-range couplings are motivated by recent experiments \cite{Nadj_et_al14,Pascual_et_al16,Menard_et_al15,Heinrich_et_al17,Menard_et_al16} and theoretical proposals \cite{Ronty_et_al_15,Li_et_al_16_2D} of lattices of magnetic impurities and islands in 2D superconductors. In addition, the imbalance between pairing and hopping decaying exponents could be achieved through Floquet driving fields \cite{Benito_et_al14}. The enhancement of topological phases by long-range couplings can be of importance for the observation and manipulation of Majorana modes without the need of a precise fine-tuning of the chemical potential.

\begin{acknowledgments}

We thank Leonid Glazman and Tharnier Oliveira for valuable conversations. M.A.MD. and O.V. thank the Spanish MINECO grant FIS2012-33152, the CAM research consortium QUITEMAD+ S2013/ICE-2801. O.V. thanks Fundaci\'on Rafael del Pino, Fundaci\'on Ram\'on Areces and RCC Harvard. The research of M.A.M.-D. has been supported in part by the U.S. Army Research Office through Grant No. W911N F-14-1-0103. The work at MIT is supported by the DOE Office of Basic Energy Sciences, Division of Materials Sciences and Engineering under Award DE-SC0010526. LF is partly supported by the David and Lucile Packard Foundation.

\end{acknowledgments}


\begin{thebibliography}{99}
\bibitem{remark} This is in contrast to the standard topological transition between different chiral phases tuning the chemical potential.
\end{thebibliography}
\end{document}